\documentclass[11pt]{article}
\usepackage{latexsym}  
\usepackage{amssymb}
\usepackage{graphicx}
\usepackage{amsmath}

\begin{document}




\begin{center}
{\Large\bf What Physics Does  The Charged Lepton \\
Mass Relation Tell Us? }\footnote{
A talk presented at ``7th Workshop on Flavour Symmetries 
and Consequences in Accelerators and Cosmology" 
(FLASY 2018).
}

\vspace{3mm}
{\bf Yoshio Koide}

 {\it Department of Physics, Osaka University, 
Toyonaka, Osaka 560-0043, Japan} \\
{\it E-mail address: koide@kuno-g.phys.sci.osaka-u.ac.jp}

\end{center}

 
\vspace{4mm}

{\Large\bf \S 0.  Prologue }

\vspace{2mm}

The story begins from a charged lepton mass 
relation\cite{Koide_82}
$$
K \equiv \frac{m_e +m_\mu+m_\tau}{\sqrt{m_e} 
+\sqrt{m_\mu} +\sqrt{m_\tau})^2} = \frac{2}{3} .
\eqno(0.1)
$$
It is well known that the formula is excellently
 satisfied by the observed charged lepton masses \cite{PDG16}
$$
K(m_{ei}^{obs}) = \frac{2}{3}\times(0.999989\pm 0.000014) .
\eqno(0.2)
$$

1982, the formula predicted  a tau lepton mass 
$$
m_\tau^{pred} = 1776.97\ {\rm MeV} ,
\eqno(0.3)
$$
by inputting the observed mass values $m_e$ and $m_\mu$.
On the other hand, the observed mass at 1982 was
$$
(m_\tau^{obs})_{old} = 1784.2 \pm 3.2 \ {\rm MeV} .
\eqno(0.4)
$$

Ten years after (1992), an  accurate value of 
$m_\tau^{obs}$ was  reported:  
$$
(m_\tau^{obs})_{new} = 17776.99_{-0.26}^{+0.29} \ {\rm MeV} .
\eqno(0.5)
$$   
The new observed value (0.5) was excellently coincident with the prediction (0.3).
Therefore, the formula suddenly attracted a great deal of attention as seen
in Fig.1.

\vspace{3mm}

\begin{figure} [h]
\begin{picture}(600,110)(0,0)
\put(90,20){\bf 1775}
\put(190,20){\bf 1780}
\put(290,20){\bf 1785}
\put(360,20){\bf MeV}
\put(100,50){\line(+1,0){260}}
\put(50,50){\bf 1992} 
\put(100,50){\line(0,+1){10}}
\put(120,50){\line(0,+1){5}}
\put(140,50){\line(0,+1){5}}
\put(160,50){\line(0,+1){5}}
\put(180,50){\line(0,+1){5}}
\put(200,50){\line(0,+1){10}}
\put(220,50){\line(0,+1){5}}
\put(240,50){\line(0,+1){5}}
\put(260,50){\line(0,+1){5}}
\put(280,50){\line(0,+1){5}}
\put(300,50){\line(0,+1){10}}
\put(320,50){\line(0,+1){5}}
\put(340,50){\line(0,+1){5}}
\put(360,50){\line(0,+1){5}}
\put(140,50){\circle*{5}}
\put(220,90){\bf Experimental value at 1982}
\put(100,80){\line(+1,0){260}}
\put(50,80){\bf 1982} 
\put(100,80){\line(0,+1){10}}
\put(120,80){\line(0,+1){5}}
\put(140,80){\line(0,+1){5}}
\put(160,80){\line(0,+1){5}}
\put(180,80){\line(0,+1){5}}
\put(200,80){\line(0,+1){10}}
\put(220,80){\line(0,+1){5}}
\put(240,80){\line(0,+1){5}}
\put(260,80){\line(0,+1){5}}
\put(280,80){\line(0,+1){5}}
\put(300,80){\line(0,+1){10}}
\put(320,80){\line(0,+1){5}}
\put(340,80){\line(0,+1){5}}
\put(360,80){\line(0,+1){5}}
\put(284,80){\circle*{5}}
\put(284,80){\thicklines \vector(1,0){60}}
\put(284,80){\thicklines \vector(-1,0){60}}
\put(50, 110){\bf Prediction} 
\put(140,110){\circle*{5}}
\put(140,110){\thicklines \line(0,-1){80}} 
\end{picture}
\end{figure}

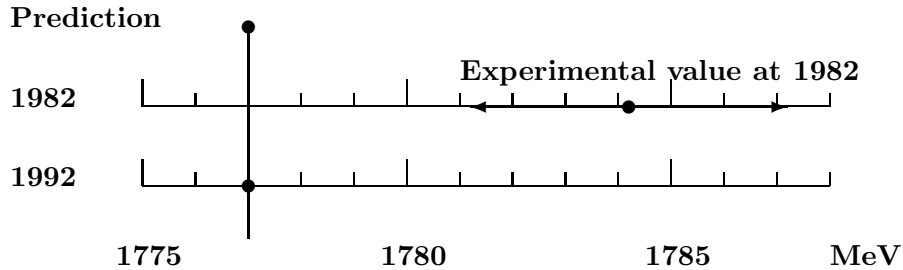
{Fig.1 \  Predicted and experimental values of the tau mass: 
The experimental error bar in 1992 is too small, so that we cannot denote 
it in the figure. }

\newpage

Thus, it is not that the formula (0.1) was proposed by taking 
the observed values into consideration.

I would like to emphasize that the formula $K=2/3$ should 
be never satisfied with the observed 
charged lepton masses in the theoretical point of view. 
This is the main motivation of my present talk.

In general, the ``mass" in the relation derived in
 a field theoretical model means the ``running" 
mass, instead of the ``pole" mass.  Therefore, 
the charged lepton mass relation $K=2/3$ should be 
never satisfied by pole masses.    Nevertheless,  the 
relation is excellently satisfied  by the pole masses 
as seen in Eq.(0.2). 
This accuracy is excellent enough to believe that 
the coincidence (0.2) is not accidental, but suggests 
a nontrivial physics behind it. 
Thus, if we take the coincidence seriously, we 
should treat  the renormalization group (RG) 
effects carefully.  This was first pointed by 
Sumino\cite{Sumino_09}.

Therefore, my talk is arranged as follows:
The present topic is not phenomenological one.
The purpose of my talk is to review of a field 
theoretical study by Sumino and recent 
development.
 
1.  Why is the excellent coincidence is so problematic? 

2.  Derivation of the mass formula

3.  Sumino mechanism

4.  Modified Sumino model 

5.  Recent development

\vspace{5mm}

{\Large\bf \S 1.  Why is the excellent coincidence so problematic? }

\vspace{2mm}

The charged lepton mass relation was derived based
 on a field theoretical model as reviewed in the next section.
Therefore, we have to use the running masses for the formula 
$K=2/3$, not the pole masses.  
However, if we use pole masses, then, we obtain
$$
K(m_{ei}^{run}) = \frac{2}{3}\times(1.00189\pm 0.00002) ,
\eqno(1.1)
$$
(at $\mu =m_Z$). 
The agreement is not so excellent. 

Are the present observed mass values mistaken?
Is the coincidence (0.2)  accidental? 
Will  a future experimental value be changed, and 
will the problem disappeared?   
However, such a case is not likely.

This is a serious theoretical problem.

\vspace{5mm}

{\Large\bf \S 2.  Derivation of the mass formula}

\vspace{2mm}

Prior to review of the Sumino model, let us review the 
derivation\cite{Koide_MPLA90} of the formula 
$K=2/3$  in briefly.  

First,  
we introduce a scalar $\Phi$ which is a nonet of 
a family symmetry U(3) and whose vacuum expected value 
(VEV) is given as
$$
\langle \Phi\rangle = v_0 {\rm diag}(z_1, z_2, z_3),
\eqno(2.1)
$$
with $z_1^2+z_2^2 +z_3^2 =1 $. 

In the model, the charged lepton mass matrix 
$M_e$ is given by   
$$
M_e = k_e \langle \Phi\rangle  \langle \Phi\rangle .
\eqno(2.2)
$$
(The structure (2.2) may be considered from a seesaw 
like mechanism.)  

Then we assume the following scalar potential: 
$$
V= \mu^2 [\Phi\Phi] + \lambda [\Phi\Phi\Phi\Phi] 
+\lambda' [\Phi_8\Phi_8][\Phi]^2 ,
\eqno(2.3)
$$
where $\Phi_8$ is an octet part of the nonet scalar $\Phi$, 
$$
\Phi_8 \equiv \Phi - \frac{1}{3} [\Phi] {\bf 1} .
\eqno(2.4)
$$
(${\bf 1}$ is a unit matrix:  ${\bf 1}=$diag$(1,1,1)$.)  
Here and hereafter, for convenience, we denote Tr$[A]$ as 
$[A]$ simply. 
Then, the condition $\partial V/\partial\Phi =0$  leads to 
$$
\frac{\partial V}{\partial\Phi} = 2 \left( \mu^2 + 
\lambda [\Phi\Phi] +\lambda' [\Phi]^2 \right) \Phi
+\lambda' \left( [\Phi\Phi] -\frac{2}{3} [\Phi]^2 \right) {\bf 1} .
\eqno(2.5)
$$
Hereafter, for convenience, we denote $\langle \Phi \rangle$ 
as $\Phi$ simply. 

We want a solution   $\Phi \neq {\bf 1}$,  so that
the coefficients of  $\Phi$  and ${\bf 1}$  must be zero.
Then, we obtain
 $$
 \mu^2 +\lambda [\Phi\Phi] +\lambda' [\Phi]^2 =0 ,
\eqno(2.6)
$$
and
$$
[\Phi\Phi] - \frac{2}{3} [\Phi]^2 = 0 .
\eqno(2.7)
$$
The relation (2.6) fixes the scale of the VEV value $\Phi$,
and the relation (2.7) gives  just our mass relation $K=2/3$. 
Note that Eq.(2.7) is independent of
 the potential parameters $\mu$ and $\lambda$. 

\vspace{2mm}

Also, recently, we have obtained another mass formula
\cite{Koide_PLB18}
$$
\kappa \equiv \frac{ {\rm det}\Phi }{[\Phi]^3} 
= \frac{ \sqrt{ m_e m_\mu m_\tau} }{ (\sqrt{m_e} + 
\sqrt{m_\mu} + \sqrt{m_\tau} )^3 } = 
\frac{1}{2\cdot 3^5} = \frac{1}{486},
\eqno(2.8)
$$
in addition to the formula $K=2/3$:
$$
K \equiv \frac{ [\Phi \Phi]}{[\Phi]^2} =
\frac{m_e +m_\mu+m_\tau}{\sqrt{m_e} +\sqrt{m_\mu} +\sqrt{m_\tau})^2}
 = \frac{2}{3} .
\eqno(2.9)
$$

Note that those relation are invariant
under a transformation
$$
(m_e, m_\mu, m_\tau) \ \ \rightarrow \ \ 
(\lambda m_e, \lambda m_\mu, \lambda m_\tau) .
\eqno(2.10)
$$
This will become important in order to understand 
the Sumino mechanism which is reviewed in the next section.

\newpage

{\Large\bf \S 3.  Sumino mechanism}

\vspace{2mm}

The deviation between  $K(m_i(\mu))$ and 
$K(m_i^{pole})$ is caused by the logarithmic term  
the QED correction\cite{Arason}
$$
m_i (\mu) = m_i^{pole} \left\{ 1 - \frac{\alpha(\mu)}{\pi} 
\left( 1 + \frac{3}{4}  \log\frac{\mu^2}{(m_i^{pole})^2} \right) 
\right\} , 
\eqno(3.1)
$$
where $m_i(\mu)$ and $m_i^{pole}$ are running mass and 
pole mass, respectively.  
(Hereafter, for simplicity, we denote $m_i^{pole}$ as $m_i$.) 
If the logarithmic term $\log m_i^2$ is absent, then 
the formula $K=2/3$ will also be satisfied by the running 
masses as we seen in Eq.(1.10). 

In 2009, Sumino\cite{Sumino_09} proposed an attractive mechanism: 
He assume U(3) family gauge bosons (FGBs) $A_i^{\ j }$ 
with their masses $(M_{ij})^2 \propto (m_i +m_j)$. 
Then, the unwelcome term $\log(m_i/\mu)^2$ in Eq.(3.1) is
canceled by the factor $\log(M_{ii}/\mu)^2$ 
in the radiative mass term due to FGB. 
(Note that in his model, only FGBs $A_i^{\ j}$ 
with $i=j$ contribute to the radiative diagram.)  

However, we should notice that the Sumino model has 
some serious shortcomings. 
In order to cancel the $\log m_i$ term in the QED 
contribution by the $\log M_{ii}$ term in the 
FGB contribution, we must 
consider an origin of the minus sign. 
Sumino has assumed that the left- and right-handed 
charged leptons $e_L$ and $e_R$ have the same sign 
coupling constants $e$ and $e$ for photon, respectively,  
but the coupling constants for FGBs takes $+g$ and $-g$
for $e_L$ and $e_R$, respectively.  In other words,  
Sumino has assigned the charged leptons $e_L$ and 
$e_R$ to ${\bf 3}$ and ${\bf 3}^*$ of U(3) family, respectively. 
Therefore, his model is not anomaly free.
Besides, in his model, unwelcome decay modes 
with $\Delta N_{family} = 2$ inevitably appear. 

\vspace{5mm}

{\Large\bf \S4.  Modified Sumino model}

\vspace{2mm}

In order to avoid these defects, Yamashita and YK  proposed 
a modified Sumino model\cite{K-Y_PLB12} with 
$(e_L, e_R) = ({\bf 3}, {\bf 3})$ of U(3) family.   
In this model, the minus sign comes from the following idea: 
The family gauge bosons have an inverted mass hierarchy, i.e.
$$
  M_{ii}^2 \propto (m_i)^{-1} .
\eqno(4.1)
$$
Then, because of $\log M_{ii}^2 \propto -\log m_i$, we 
can obtain the minus sign for the cancellation 
without taking  $(e_L, e_R) = ({\bf 3}, {\bf 3}^*)$.

In the modified assignment 
$(e_L, e_R) = ({\bf 3}, {\bf 3})$ with the 
inverted mass hierarchy  (4.1),  
FGB with the lowest mass is $A_3^{\ 3}$. 
Note that the family number $i=1,2,3$ is 
defined by the charged lepton sector, i.e. 
$i=(1,2,3) = (e, \mu, \tau)$. 
On the other hand, for the quark sector, 
we do not have any constraint from 
experimental observations. 
Both cases $(u_1, u_2, u_3)= (u, c, t)$
and $(u_1, u_2, u_3)= (t, c, u)$ are allowed.
If we choose the latter case, we can expect 
FGBs with considerably lower masses,
 e.g. we can suppose $M(A_3^{\ 3}) \sim$ a few TeV, 
 because the inverted family number assignment 
for quarks  weakens severe constraints from 
$K^0$-$\bar{K}^0$ and   $D^0$-$\bar{D}^0$ mixing data, 
so that we can obtain  considerably low FGB masses 
\cite{Koide_PLB14}.  
Thus, in the modified Sumino model, we can 
expect fruitful phenomenology.  
 For examples, 
 see Ref.\cite{K-Y_PLB16} for $\mu$ -$e$ conversion, 
and see Ref.\cite{ K-Y-Y_PLB15} for $A_1^{\ 1}$ production at LHC.  
However, note that in our model the transition
$\mu \rightarrow  e + \gamma$ is exactly forbidden.


\vspace{5mm}

{\large\bf 5.  Recent development and summary}

\vspace{2mm}

There is another effect which disturbs the $K$-relation:
 $\Phi_8 \leftrightarrow \Phi_0 \equiv [\Phi]/\sqrt{3}$ 
mixing due to renormalization effect  \cite{Yamashita_18}.  
The $K$ and $\kappa$ relations were derived from potential 
model under a non-SUSY scenario.  
Recall that there is no vertex correction in a SUSY
 model.   Therefore, if we derive the relations on the basis 
of SUSY scenario, then the problem will disappear. 
Very recently, we succeeded to  re-derive the $K$ and 
$\kappa$ relations on the basis of SUSY scenario 
\cite{K-Y_18}. .  
 Thus,  we can understand why the $K$- and $\kappa$-relations 
can keep the original forms.

In conclusion, we have discussed why the $K$ relation is so
beautifully satisfied by the pole masses, 
not the running masses. 
Now we can understand the reason 
according to the Sumino's idea and 
the modified Sumino model. 

\vspace{5mm}



\end{document}